\documentclass[aps,superscriptaddress,twocolumn,showpacs,preprintnumbers,amsmath,amssymb]{revtex4}
\usepackage[cp1251]{inputenc}
\usepackage{amssymb,amsmath}
\usepackage[mathscr]{eucal}
\usepackage{graphicx}
\usepackage{longtable}
\usepackage[dvips]{hyperref} 
\begin{document}
\title{Quantum electrodynamic equations for magnetic resonance- and optical spectroscopic transitions}
\author{D.Yearchuck}
\affiliation{Minsk State Higher College, Uborevich Str.77, Minsk, 220096, yearchuck@gmail.com} 
\author{Y.Yerchak}
\affiliation{Byelarussian State University, Nezavisimost Avenue4, Minsk, 220030, jarchak@gmail.com}
\author{A.Alexandrov}
\affiliation{M.V.Lomonosov Moscow State University, Moscow, 119899, alex@ph-elec.phys.msu.su}	
\date{\today}
\begin{abstract}
Quantum electrodynamic equations for magnetic resonance- and optical spectroscopic transitions have been for the first time obtained. New phenomena - stochastic electrical and magnetic spin wave resonances are predicted to be the effects of EM-field quantization.
\end{abstract}
\maketitle Spectroscopic description of the systems, interacting with electromagnetic (EM) field including the description of transitional effects,
for instance, Rabi oscillations, free induction, 
spin echo in magnetic resonance spectroscopy and
corresponding optical analogues in optical spectroscopy
is achieved within the framework of gyroscopic model
based on classical Landau-Lifshitz (L-L) equation \cite{Landau}. The gyroscopic
model for transition dynamics and for description
of transitional effects was introduced formally \cite{Abragam} by
F.Bloch. Then L-L equation and Bloch equations, based on L-L equation, were substantiated quantum-mechanically
in magnetic resonance theory, however the only for
the motion of magnetic moment in external magnetic
field \cite{Abragam}, \cite{Slichter}. The application of given equations for the description of dynamics 
of magnetic resonance transitions was in fact postulated. The development of gyroscopic
model for the characterization of optical systems, interacting with EM-field was done 
 on the base of density operator formalism, see for instance \cite {Apanasevitch}, \cite {Scully} and optical semiclassical Bloch
equations were obtained. However, it will readily be observed, that there is some vagueness with clear physical meaning of the quantities, used in optical Bloch
equations. In a certain sense the vectors $\vec {P}$, $\vec {E}$ (see eq.2 below) entering
optical Bloch equations are in fact some mathematical
abstractions. Really, like
to magnetic moment and magnetic field vectors in magnetic resonance Bloch equations, both the vectors $\vec {P}$, $\vec {E}$ have necessary axial symmetry, however in artificial way \cite{Macomber}. For instance, $\vec {P}$ (it is called usually Bloch vector) has two components $P_{x}$, $P_{y}$ of polar electric dipole moment vector, that is both the components remain without any physical meaning change. At the same time
 the third component of $\vec {P}$ has entirely different physical meaning and it is considered to be proportional to the population difference between the energy levels, see for example \cite {Apanasevitch}, \cite{Scully, Macomber}. Mathematically the objects, which are like to Bloch vector, can, naturally, exist, although they can produce the only affine space, which is not metrizable. However, physically the model, in which the part of components of characteristic vector possess by different symmetry (polar in given case) in comparison with the final symmetry (axial) of vector itself seems to be incorrect. Given conclusion is especially evident, if to take into account,
that the nature of the spectroscopic transitions has
to be the same in optics and in magnetic resonance, since it is experimentally confirmed, that they can be described by in fact the same mathematical equation. Let us remember, that the analogue of optical Bloch vector in magnetic resonance spectroscopy is full magnetic moment, that is pure axial vector with peer all the three components. Second aspect, concerning known variants of gyroscopic model, is the fact, that all known versions are classical or semiclassical. In other words, EM field is considered always classically. However, it is well known, that some phenomena, for example quantum beats \cite{Scully}, cannot be described classically or semiclassically in principle. Therefore, it is reasonable to suggest, that with help of complete quantum model can be predicted some new  effects, accompanying magnetic resonance and optical absorption phenomena.

The main aim of present work is the derivation of fully-quantum equations for the spectroscopic transitions in the system EM field plus matter with clear physical sense of all the quantities (for the case of simple 1D model of quantum-mechanical matter subsystem) 
and to predict some new effects, which can appear due to EM field quantization.

Classical L-L equation \cite{Landau} is given by the following equation (\ref{eq1}) 
\begin{equation}
\label{eq1}
\frac{d\vec {S}}{dt}=[\gamma _{_H} \vec {H}\times \vec {S}], 
\end{equation}
where $\vec {S}$ is magnetic moment, $\vec {H}$ is effective magnetic field, 
$\gamma _{H} $ is gyromagnetic ratio. Optical analogue of L-L equation is the equation (\ref{eq2}),
 \begin{equation}
\label{eq2}
\frac{d\vec {P}}{dt}=[\vec {P}\times \gamma _{_E} \vec {E}], 
\end{equation}
where $\vec {P}$ is Bloch vector, $\vec {E}$ is vector, which is considered like to Bloch vector to 
be some mathematical affine space abstraction, since $E_{x}$, $E_{y}$ components 
characterize the genuine electromagnetic properties of the system and they are the components of polar vector, at the same time the axial vector component $E_{z}$ does not refer to electromagnetic characteristics. Further,
$\gamma _{E}$ is gyroelectric ratio, although it was called so the only tentatively, and its analogy with gyromagnetic ratio was 
 suggested. 
Let us consider the system, representing 
itself the periodical ferroelectrically (ferromagnetically) ordered chain of 
$n$ equivalent elements (elementary 
units), interacting with external oscillating electromagnetic 
field. It is assumed, that the interaction between elements of the chain can be described by the Hamiltonian of quantum X Y Z 
Heisenberg model in the case of a chain of magnetic dipoles and by 
corresponding optical analogue of given Heisenberg model in the case of a 
chain of electric dipoles. We will consider for the simplicity the case of 
isotropic exchange. Each elementary unit of the chain is proposed to be 
two-level quantum subsystem like to one-electron atom. Then Hamiltonians for the 
chain of electrical dipole moments and for the chain of magnetic dipole 
moments will be mathematically equivalent. The chain for distinctness of 
electrical dipole moments can be described by the following Hamiltonian
\begin{equation}
\label{eq3}
\mathcal{\hat H} = \mathcal{\hat H}^C + \mathcal{\hat H}^F + \mathcal{\hat H}^{C F},
\end{equation} 
where 
${\mathcal{\hat H}^C}$ is chain Hamiltonian by the absence of the interaction with EM field, ${\mathcal{\hat H}^F}$ is field Hamiltonian, ${\mathcal{\hat H}^{C F}}$ is Hamiltonian, describing the interaction between quantized EM field and atomic chain. In its turn, Hamiltonian ${\mathcal{\hat H}^C}$ is
\begin{equation}
\label{eq4}
{\mathcal{\hat H}^C} = {\mathcal{\hat H}^0} + {\mathcal{\hat H}^J}, 
\end{equation} 
where ${\mathcal{\hat H}^0}$ is chain Hamiltonian in the absence of the interaction between structural elementary units of the chain. We suggest, that given structural elementary units are optical (magnetic) centers with one electric (magnetic) dipole pro center. ${\mathcal{\hat H}^J}$ is Hamiltonian of exchange interaction between given centers. To obtain the equation, describing the transition dynamics, we will use the transition operator method. Given method was used in atomic spectroscopy, see, for instance, \cite{Scully}. The essence of the method, adapted for our task, is the following. Let $\left\{ {\left|m_l \right\rangle}\right\}$, $l = \overline{1,n}$, $m = \alpha, \beta$, be full set of energy states of $l-\textit{th}$ optical (magnetic) center of the chain in the absence of the interaction both between the centers themselves and with EM field. Here $\left| {\alpha _l} \right\rangle $ is lower state and $\left|{\beta _l } \right\rangle $ is upper state. They are eigenstates of $l-\textit{th}$ item in ${\mathcal{\hat H}^0}$, $l = \overline{1,n}$. It means, that ${\mathcal{\hat H}^0}$ is
\begin{equation}
\label{eq5}
\mathcal{\hat H}^0 = \sum\limits_{v=1}^n \sum\limits_m \mathcal{E}_{mv}{\left|m_v \right\rangle} {\left\langle m_v \right|}.
\end{equation}
Here $m = \alpha, \beta$, $\mathcal{E}_{mv}$ are eigenvalues of ${\mathcal{\hat H}^0}$, which correspond to the states ${\left|m_v \right\rangle}$ of $v-\textit{th}$ chain center. 
Let set up in correspondence to the states ${\left|j_v \right\rangle}$, $v = \overline{1,n}$, the operators 
\begin{equation}
\label{eq6}
{\hat\sigma_v}^{jm} \equiv {\left|j_v \right\rangle} {\left\langle m_v \right|}, 
\end{equation}
where $j = \alpha, \beta, m = \alpha, \beta $.
Then, it is evident, that operators $\hat {\sigma }_v^+ = \left| {\beta _v } \right\rangle 
\left\langle {\alpha _v } \right|$ are the operators, which transform eigenstates $\left| {\alpha_v } \right\rangle$ into $\left| {\beta_v }\right\rangle$ and operators $\hat {\sigma }_v^- = \left| {\alpha_v } \right\rangle \left\langle {\beta _v} \right|$ are the operators, transforming eigenstates $\left| {\beta_v } \right\rangle$ into $\left| {\alpha_v }\right\rangle$ for $\forall v$, $v = \overline{1,n}$. Let define also the operators $\hat {\sigma }_v^z = \left| {\beta _v } \right\rangle \left\langle {\beta _v 
} \right| - \left| {\alpha _v } \right\rangle \left\langle 
{\alpha _v } \right|$. The operators $\hat {\sigma }_v^z$ transform for $\forall v$, $v = \overline{1,n}$, the sum of states $\left| {\beta_v } \right\rangle$ + $\left| {\alpha_v } \right\rangle$ into their difference and the reverse is also true. The above defined set of spectroscopic transition operators can be completed by unit operators $\hat {\sigma }_v^ E$, $v = \overline{1,n}$, which, owing to completeness of state sets can be represented by the following relationship
\begin{equation}
\label{eq7}
\hat {\sigma }_v^E = \sum\limits_m \left| {m_v } \right\rangle \left\langle {m_v} \right|,
\end{equation}
 $m = \alpha, \beta $, 
and by zeroth operators $\hat {\sigma}_v^ 0$, $v = \overline{1,n}$, which can be defined in the following way
\begin{equation}
\label{eq8}
\hat {\sigma }_v^0 = \hat {\sigma }_v^j - \hat {\sigma }_v^j,
\end{equation}
where $j = -, +, z, E$.
It is easy to show, that at fixed $v$ the set of transition operators is closed relatively the algebraic operations of commutation and anticommutation and relatively the operation of Hermitian conjugation. For instance, the relationships for commutation rules are
\begin{equation}
\label{eq9a}
[\hat {\sigma}_v^{lm}, \hat {\sigma}_v^{pq}] = \hat {\sigma }_v^{lq} \delta_{mp} - \hat {\sigma }_v^{pm}\delta_{ql}, 
\end{equation}
that is, 
\begin{equation}
\label{eq9b}
[\hat {\sigma}_v^{-}, \hat {\sigma}_v^{z}] = 2 \hat {\sigma }_v^{-}, 
[\hat {\sigma}_v^z, \hat {\sigma }_v^{+}] = 2 \hat {\sigma }_v^+, 
[\hat {\sigma}_v^{+}, \hat {\sigma }_v^{-}] = \hat {\sigma }_v^z. 
\end{equation}
 So it is seen, that in comparison with the method of density matrix, which gives the relationship for probabilities of the states, we are dealing in spectroscopic transition operators' method immediately with the states. 
Let us define the vector operator: 
\begin{equation}
\label{eq10}
\hat {\vec {\sigma }}_v =\hat {\sigma }_v^- \,\vec {e}_+ +\hat {\sigma 
}_v^+ \,\vec {e}_- +\hat {\sigma }_v^z \,\vec {e}_z .
\end{equation}
It seems to be the most substantial for the subsequent consideration, that a set 
of $\hat {\sigma }_v^m $ operators, where m is z, +, -, 0, E, produces algebra, 
which is isomorphic to $S = 1/2$ Pauli matrix algebra, completed by unit and zeroth matrices, that is, 
mappings $f_v :\,\hat {\sigma }_v^m \to \sigma _P^m $ realize 
isomorphism. Here $v$ is a number of chain unit, $\sigma _P^m $ 
is extended set of Pauli matrices for the spin of $1/2$. Consequently, from 
physical point of view $\hat {\vec {\sigma }}_v$ represents itself some 
vector operator, which is proportional to vector operator of the spin of $v-\textit{th}$ chain 
unit. Vector operators $\hat {\vec {\sigma }}_v$ produce also linear space over field of complex numbers, which 
can be called transition space. It is 3-dimensional (in the case of two-level systems), that is 3 matter operator 
equations of the motion for components of $\hat {\vec {\sigma }}_v$ is necessary for correct 
description of optical transitions in semiclassical approach. 
Isomorphism of Pauli matrix algebra and spectroscopic transition operators' algebra allows to change the spin variables in Heisenberg exchange Hamiltonian by components of vector $\hat {\vec {\sigma }}_v$. Then Hamiltonian ${\mathcal{\hat H}^C}$, taking into account 
the relationship (\ref{eq5}), can be represented in the form
\begin{equation}
\label{eq10a}
\begin{split}
\raisetag{40pt}
&\mathcal{\hat H^C} = \frac{\hbar}{2} \sum\limits_{v = 1}^n \omega _v {\hat {\sigma}_v^z } + 
\frac{1}{2}\sum\limits_{v = 1}^n (\mathcal{E}_{v\alpha} + \mathcal{E}_{v\beta} ){\hat {\sigma }_v}^{E} \\
& \ \  + \sum\limits_{v = 1}^n [J_{\vec {E}} (\hat {\sigma} _v^ + \hat {\sigma} _{v + 1}^- + \hat {\sigma} _v^ -  \hat {\sigma} _{v + 1}^ + + \frac{1}{2}\hat {\sigma} _v^z \hat {\sigma} _{v + 1}^z ) + H.c.],
\end{split}
\end{equation}
where $J_{\vec E}$ is optical analogue of the 
exchange interaction constant. Here, in correspondence with the suggestion, 
$J_{\vec E} =J_{\vec E}^x =J_{\vec E}^y =J_{\vec E}^z $. The first and the second items in (\ref{eq10a}) were obtained taking into consideration the only definition of operators $\hat {\sigma }_v^z$, $v = \overline{1,n}$. The $\textit{v-th}$ component of the sum in the second item of (\ref{eq10a}) represents physically the operator of the shift of the energy reference for $\textit{v-th}$ chain unit. Expression (\ref{eq10a}) is obtained by using of the basis 
$\vec {e}_+ = \frac{1}{2} (\vec {e}_x +i\vec {e}_y ), \,\vec {e}_- = \frac{1}{2} (\vec {e}_x -i\vec{e}_y ), \vec {e}_z = \vec {e}_z$. Note, that in given basis the representation of components of spin vector operator by means of Pauli matrices and, correspondingly, the representation of components of spectroscopic transition vector operator $\hat {\vec {\sigma }}_v$ by Pauli matrices, has the most simple form. We also suggest 
 that the full set of eigenstates for all the $n$ elements - $\left\{ {\left| {\alpha _v } \right\rangle}\right\}$ and $\left\{ {\left| {\beta _v } \right\rangle}\right\}$ - can also be considered to be basis set in Hilbert space for Hamiltonian $\mathcal{\hat H}$. It is evident, that given assumption can be realized strictly the 
only by the absence of the interaction between the elements. At the same 
time proposed model will rather well describe the real case, if the 
interaction energy of adjacent elements is much less of the splitting energy $\hbar \omega _v =\mathcal{E}_{v\beta} -\mathcal{E}_{v\alpha}$ between the energy levels, 
corresponding to the states $\left|\alpha_v\right\rangle$ and $\left|\beta_v\right\rangle$. Given case includes in fact all known 
experimental situations. In the case of the 
chain of magnetic dipole moments $J_{\vec E}$ is replaced by the exchange interaction constant 
$J_{\vec H}$, and the 
frequency $\omega_v $ is replaced by $ \frac{1}{\hbar}g_{vH} \beta_{H} H_0 = \gamma_{vH} H_0 $, where $H_0$ is external static magnetic field, $\beta_{H}$ is Bohr magneton, $g_{vH}$ is $g$-tensor, which is assumed for the simplicity to be isotropic. It is clear, that Hamiltonian $\mathcal{\hat H}^{C F}$ of interaction of quantized EM field with atomic chains can also be represented in the set of variables, which includes the components of spectroscopic transition vector operator $\hat {\vec {\sigma }}_v$. Really, suggesting dipole approximation to be true and polarization of field components to be fixed, we have 
\begin{equation}
\label{eq11}
\begin{split}
\raisetag{40pt}
\mathcal{\hat H}^{C F} = -\sum\limits_{j = 1}^n \sum\limits_{l \neq m} \sum\limits_{m} \sum\limits_{\vec k} [ p_j^{lm} \hat {\sigma}_j^{lm} (\vec e_{\vec k} \vec e_ {\vec P_j}) \mathfrak{E}_{\vec k} \hat{a}_{\vec k} \times \\
e^{ - i \omega_{\vec k} t+ i \vec k \vec r} + H.c. ], 
\end{split}
\end{equation}
where $p_j^{lm}$ is matrix element of operator of electric dipole moment $\vec P_j$ of $\textit{j-th}$ chain unit between the states $\left| {l_j} \right\rangle$ and 
$\left| {m_j} \right\rangle$ with $l_j = \alpha_j, \beta_j, m_j = \alpha_j, \beta_j$, $\vec e_{\vec k}$ is unit polarization vector, $\vec e_{\vec P_j}$ is unit vector along $\vec P_j$-direction, $\mathfrak{E}_{\vec k}$ is the quantity, which has the dimension of electric field strength, $\vec k$ is wave vector, $\hat{a}_{\vec k}$ is field annihilation operator. In the suggestion, that the contribution of spontaneous emission is relatively small, we will have $p_j^{lm} = p_j^{ml} \equiv p_j $, where $l = \alpha, \beta, m = \alpha, \beta$. Let define the function
\begin{equation}
\label{eq11a}
 q_{j \vec k} = - \frac{1}{\hbar} p_j (\vec e_{\vec k} \cdot \vec e_{\vec P_j}) \mathfrak{E}_{\vec k} e^{ - i \omega_{\vec k} t+ i \vec k \vec r}
\end{equation}
Then the expression (\ref{eq11}) can be rewritten in the form
\begin{equation}
\label{eq11b}
\mathcal{\hat H}^{C F} = \sum\limits_{v = 1}^n \sum\limits_{\vec k} [q_{j \vec k} (\hat {\sigma}_j^- + \hat {\sigma}_j^+) \hat{a}_{\vec k} + (\hat {\sigma}_j^- + \hat {\sigma}_j^+) \hat{a}_{\vec k}^{ +} {q^*}_{j \vec k}],
\end{equation}
where $\hat{a}_{\vec k}^{ +}$ is creation operator, superscript $*$ in ${q^*}_{j \vec k}$ means complex conjugation. Field Hamiltonian has usual form
\begin{equation}
\label{eq11c}
\mathcal{\hat H}^{F} = \sum\limits_{\vec k} \hbar \omega_{\vec k} (\hat{a}_{\vec k}^{ +} \hat{a}_{\vec k} + \frac{1}{2}).
\end{equation}
The equations of the motion for spectroscopic transition operators $\hat {\vec {\sigma }}_l$ and for field operators $\hat{a}_{\vec k}$, $\hat{a}_{\vec k}^{ +}$ are
\begin{equation}
\label{eq12} 
i \hbar \frac{\partial\hat {\vec\sigma}_l} {\partial t} = [\hat {\sigma}_l^{-}, \mathcal{\hat H}] \vec e_+ + [\hat {\sigma}_l^{+}, \mathcal{\hat H}] \vec e_{-} + [\hat {\sigma} _l^{z}, \mathcal{\hat H}] \vec e_z, 
\end{equation}
\begin{equation}
\label{eq12a}
i \hbar \frac{\partial\hat{a}_{\vec k^{'}} }{\partial t} = [\hat{a}_{\vec k^{'}}, \mathcal{\hat H}], 
i \hbar \frac{\partial\hat{a}_{\vec k^{'}}^+}{\partial t} = [\hat{a}_{\vec k^{'}}^{ +}, \mathcal{\hat H}].
\end{equation}
Then, using (\ref{eq3}, \ref{eq10}, \ref{eq11b}, \ref{eq11c}, \ref{eq9b}) 
we obtain
\begin{subequations}
\label{eq13}
\begin{gather}
\frac{\partial \hat\sigma_l^z}{\partial t} = \frac{2 i}{\hbar} (\hat\sigma_l^{+} - \hat\sigma_l^{-}) \sum\limits_{\vec k} (q_{l \vec k} \hat{a}_{\vec k} + \hat{a}_{\vec k}^{+} {q^*}_{l \vec k}) 
\\ 
+ \frac{{2iJ_{\vec E} }}{\hbar }\left(\left\{ {\hat\sigma _l^ - } \right.,\left. {(\hat\sigma _{l + 1}^ + + \hat\sigma _{l - 1}^+ )} \right\} - 
\left\{ {\hat\sigma _l^ + } \right.,\left. {(\hat\sigma _{l + 1}^ - + \hat\sigma _{l - 1}^ - )} \right\}\right),\nonumber
\\
\frac{\partial \hat\sigma_l^+}{\partial t} = i \omega_l \hat\sigma_l^{+} - \frac{i}{\hbar} \hat\sigma_l^{z} \sum\limits_{\vec k} (q_{l \vec k} \hat{a}_{\vec k} + \hat{a}_{\vec k}^{+} {q^*}_{l \vec k})
\\
+\frac{{iJ_{\vec E} }}{\hbar } \left(\left\{ {\hat\sigma _l^ + } \right.,\left. {(\hat\sigma _{l + 1}^z  + \hat\sigma _{l - 1}^z )} \right\} - \left\{ {\hat\sigma _l^z } \right.,\left. {(\hat\sigma _{l + 1}^ +  + \hat\sigma _{l - 1}^ +  )} \right\}\right),\nonumber
\\ 
\frac{\partial \hat\sigma_l^-}{\partial t} = -i \omega_l \hat\sigma_l^{-} + \frac{i}{\hbar} \hat\sigma_l^{z} \sum\limits_{\vec k} (q_{l \vec k} \hat{a}_{\vec k} + \hat{a}_{\vec k}^{+} {q^*}_{l \vec k}) 
\\ 
+ \frac{{iJ_{\vec E} }}{\hbar } \left(\left\{ {\hat\sigma _l^z } \right.,\left. {(\hat\sigma _{l + 1}^ - + \hat\sigma _{l - 1}^ - )} \right\} - \left\{ {\hat\sigma _l^ - } \right.,\left. {(\hat\sigma _{l + 1}^z  + \hat\sigma _{l - 1}^z )} \right\}\right), \nonumber
\end{gather}
\end{subequations}
\begin{equation}
\label{eq14a}
\frac{\partial \hat{a}_{\vec k^{'}} }{\partial t} = -i \omega_{\vec k^{'}} \hat{a}_{\vec k^{'}} - \frac{i}{\hbar} \sum\limits_{j = 1}^n (\hat\sigma_j^{+} + \hat\sigma_j^{-}) {q^*}_{j \vec k^{'}},
\end{equation}
\begin{equation}
\label{eq14b}\frac{\partial \hat{a}_{\vec k^{'} }^+}{\partial t} = i \omega_{\vec k^{'}} \hat{a}_{\vec k^{'} }^+ + \frac{i}{\hbar} \sum\limits_{j = 1}^n (\hat\sigma_j^{+} + \hat\sigma_j^{-}) {q}_{j \vec k^{'}},
\end{equation}
where expressions in braces $\left\{ {\,,\,} \right\}$ are anticommutators.
Let define vector operators
\begin{equation}
\label{eq15b}
\hat {\vec {\mathcal{G}}}_{l - 1,l + 1} = \hat {\mathcal{G}}_{l - 1,l + 1}^-  \vec e_ +  + \hat {\mathcal{G}}_{l - 1,l + 1}^ +  \vec e_ - + \hat {\mathcal{G}}_{l - 1,l + 1}^z \vec e_z,
\end{equation}
 where their components are
\begin{subequations}
\label{eq15}
\begin{gather}
\hat {\mathcal{G}}_{l - 1,l + 1}^-  = -\frac{1 }{\hbar} \sum\limits_{\vec k}\hat{b}_{l \vec k} - \frac{{J_{\vec E} }}{\hbar }(\hat\sigma _{l + 1}^- + \hat\sigma _{l - 1}^-), \\
\hat {\mathcal{G}}_{l - 1,l + 1}^+ = -\frac{1}{\hbar} \sum\limits_{\vec k}\hat{b}_{l \vec k} - \frac{{J_{\vec E} }}{\hbar }(\hat\sigma _{l + 1}^+ + \hat\sigma _{l - 1}^ + ), \\
\hat {\mathcal{G}}_{l - 1,l + 1}^z = - \omega_{l} - \frac{{J_{_E} }}{\hbar }(\hat\sigma _{l + 1}^z + \hat\sigma _{l - 1}^z ).
\end{gather}
\end{subequations}
Here $\hat{b}_{l \vec k} = q_{l \vec k} \hat{a}_{\vec k} + \hat{a}_{\vec k}^{+} {q^*}_{l \vec k}$.
Then the equations (\ref{eq13}, \ref{eq14a}, \ref{eq14b}) can be represented in compact vector 
form
\begin{equation}
\label{eq15c}
\frac{{\partial \hat {\vec {\sigma}} _l }}{{\partial t}} = 2 \left\| g \right\| \left[ {\hat {\vec {\sigma}}_l \otimes \hat {\vec {\mathcal{G}}}_{l - 1,l + 1} } \right],
\end{equation}
\begin{equation}
\label{eq15d}
\begin{split}
\raisetag{40pt}
\frac{\partial}{\partial t} 
\left[
\begin{array}{*{20}c}
 {\hat{a}_{\vec k^{'}}} \\
 \\
 {\hat{a}_{\vec k^{'}}^+} \\
\end{array} 
\right] = -i \omega_{\vec k^{'}} \sigma_P^z \left[
\begin{array}{*{20}c}
 {\hat{a}_{\vec k^{'}}} \\
 \\ 
 {\hat{a}_{\vec k^{'}}^+} \\
\end{array} 
\right] \\ + \frac{i}{\hbar}
\left[
\begin{array}{*{20}c}
{-\sum\limits_{j = 1}^n (\hat\sigma_j^{+} + \hat\sigma_j^{-}) {q^*}_{j \vec k^{'}}} \\
{\sum\limits_{j = 1}^n (\hat\sigma_j^{+} + \hat\sigma_j^{-}) {q}_{j \vec k^{'}}} \\
\end{array} \right],
\end{split}
\end{equation}
where $l = \overline{2,n-1}$, $\left\| g \right\|$ is
\begin{equation}
\label{eq15e}
\left\| g \right\| = 
\left[
\begin{array}{*{20}c}
 {1} & {0} & {0} \\
 {0} & {1} & {0} \\
 {0} & {0} & {4} \\
\end{array} 
\right], 
\end{equation} $\sigma_P^z $ is Pauli z-matrix.
It is remarkable, that vector product of vector operators in (\ref{eq15c}) can be 
calculated by using of known expression (\ref{eq16}) with additional coefficient ${\frac{1}{2}}$ the only, which is appeared, since
\begin{equation}
\label{eq16}
\left[ {\hat {\vec {\sigma}} _l \otimes \hat {\vec {\mathcal{G}}}_{l - 1,l + 1} } \right] = \frac{1}{2} \left| {\begin{array}{*{20}c}
 {\vec e_- \times \vec e_z} & {\hat{\sigma}_l^-} & {\hat {\mathcal{G}}_{\,\,l - 1,l + 1}^-} \\
 {\vec e_z \times \vec e_+} & {\hat{\sigma}_l^+} & {\hat {\mathcal{G}}_{\,\,l - 1,l + 1}^+} \\
 {\vec e_+ \times \vec e_-} & {\hat{\sigma}_l^z} & {\hat {\mathcal{G}}_{\,\,l - 1,l + 1}^z} \\
\end{array}} \right|',
\end{equation}
the products of two components of two vector operators are changed by anticommutators of corresponding 
components. Given detail is mapped by symbol $\otimes$ in (\ref{eq15c}) and by symbol $'$ in determinant (\ref{eq16}).
The definition proposed seems to be natural generalization of vector product for 
the case of operator vectors. Really, the result will 
 be independent in given case on a sequence of components of both the operator vectors in their 
products like to vector product for usual vectors. Naturally, the expression for determinant in (\ref{eq16}) can be used for calculation of vector product of usual 
vectors too. 
Taking into account the physical sense of vector operators $\hat 
{\vec {\sigma }}_l$ we conclude, that equations (\ref{eq15c}, \ref{eq15d}) represent themselves 
required quantum electrodynamic difference-differential equations (the time is 
varied continuously, the coordinates are varied discretely) for the 
description of the dynamics of the spectroscopic transitions (in the frames 
of the model proposed). It follows from here in view of isomorphism of algebras of 
operators $\hat {\vec {\sigma }}_k $ and components of the spin, 
that the (\ref{eq15c}) is quantum electrodynamic equivalent to L-L equation (in its 
difference-differential form). Consequently we have proved the possibility 
to use L-L equation for the description of the dynamics of spectroscopic 
transitions, and, consequently, for the description of transitional effects. In comparison with semiclassical description, where the description of dynamics of spectroscopic 
transitions is exhausted by one vector equation (L-L equation or L-L based equation), in the case of completely quantum consideration L-L type equation describes the only one subsystem of two-part field-matter system. Consequently, we can really expect new effects, which can be predicted the only by QED consideration of resonance transition phenomena. Further some qualitative reasoning are proposed in favor of given conclusion. It can be shown, that the equations (\ref{eq15c}, \ref{eq15d}) represent themselves vector-operator difference-differential generalization of the system, which belong to well known family of equation systems - Volterra model systems, widely used in biological tasks of population dynamics studies, which in its turn is generalization of Verhulst equation. So, the interest represents for instance the fact, that by some parameters in two-sybsystem Volterra model the stochastic component in solution is appeared.
It means, that the results obtained allow to predict, for instance, 
stochastic electric (ESWR) and magnetic (MSWR) spin wave resonances, that is spin wave resonances, where stochastic component for the quantities (all or some), characterising the resonances can be appeared. It will also be new stride toward widening EWR.-state frontier in possible practical applications. For instance, taking into account the same representation (to constant number factor) of spin vector operator and vector operator of spectroscopic transitions (\ref{eq10}) we can detect,
 along with magnetic resonance methods, a spin value of 
particles, quasiparticles, impurities or other centers in solids by optical 
methods, that is by the study of ESWR or transitional optical analogues of magnetic resonance phenomena.

Let us discuss some dephasing and relaxation processes, which can take place by spectroscopic transitions. Substantial role in dephasing and relaxation processes in solid state systems plays electron-phonon interaction, see for instance \cite{Zhu_05, WilsonRae_02}. To take electron-phonon interaction into consideration, we add to Hamiltonian \eqref{eq3} the following terms

\begin{equation}
\label{h_phonon}
    \mathcal{\hat H}^{P} = \sum\limits_{\vec q} \hbar \nu_{\vec q} (\hat{b}_{\vec q}^{ +} \hat{b}_{\vec q} + \frac{1}{2}),
\end{equation}
\begin{equation}
\label{h_e_phonon}
    \mathcal{\hat H}^{CP} =\sum\limits_{j=1}^n \sum\limits_{\vec q}  \lambda_{\vec q} (\hat{b}_{\vec q}^{ +} +\hat{b}_{\vec q})\hat{\sigma}^z_j,
\end{equation}
where $\hat{b}_{\vec q}^+$ ($\hat{b}_{\vec q}$) is the creation
(annihilation) operator of the phonon with momentum ${\vec q}$ and
energy $\hbar \nu_{\vec q} $, $\lambda_{\vec q}$ is electron-phonon coupling constant. From equations of motion \eqref{eq12}, \eqref{eq12a} we obtain the additional terms in equations for $\hat{\sigma}^{\pm}_l$: $-2\sum_{\vec q}  \lambda_{\vec q} (\hat{b}_{\vec q}^{ +} +\hat{b}_{\vec q})\hat{\sigma}^{\pm}_l$. Equations for $\hat{\sigma}^{z}_l$,  $\hat{a}_{\vec k^{'}}$, $\hat{a}_{\vec k^{'}}^+$ do not change. We obtained also the equations for new variables $\hat{b}_{\vec q}^{ +}$, $\hat{b}_{\vec q}$:
\begin{equation}
\label{eq_b}
\frac{\partial \hat{b}_{\vec q} }{\partial t} = -i \nu_{\vec q} \hat{b}_{\vec q} - \frac{i}{\hbar} \sum\limits_{j = 1}^n \lambda_{\vec q}\hat\sigma_j^{z},
\end{equation}
\begin{equation}
\frac{\partial \hat{b}_{\vec q}^+}{\partial t} = i \nu_{\vec q} \hat{b}_{\vec q}^+ + \frac{i}{\hbar} \sum\limits_{j = 1}^n \lambda_{\vec q}\hat\sigma_j^{z}.
\end{equation}
Given equations can be formally integrated to yield 
\begin{equation}
    \label{b(t)}
\hat{b}_{\vec q}(t) = \hat{b}_{\vec q}(0)e^{-i\nu_{\vec q}t} - \frac{i}{\hbar} \sum\limits_{j = 1}^n \lambda_{\vec q}\int\limits_0^t \hat\sigma_j^{z}(t')e^{-i\nu_{\vec q}(t-t')}dt'
\end{equation}
and analogously for $\hat{b}_{\vec q}^+(t)$. Hence the additional terms in equations for $\hat{\sigma}^{\pm}_l$ take the form

\begin{equation}
    \label{add}
\begin{split}
-2 \sum_{\vec q} \lambda_{\vec q}[\hat{b}_{\vec q}(0)e^{-i\nu_{\vec q}t} +\hat{b}_{\vec q}^+(0)e^{i\nu_{\vec q}t}]\hat{\sigma}^{\pm}_l(t)\\
 + \frac{4}{\hbar} \sum_{\vec q}\sum\limits_{j = 1}^n \lambda_{\vec q}\int\limits_0^t \hat\sigma_j^{z}(t')\sin[\nu_{\vec q}(t-t')]dt'\hat{\sigma}^{\pm}_l(t).
\end{split}
\end{equation}

Thus, QED equations for dynamics of spectroscopic transitions have been for the first time derived with clear physical sense of the quantities for both radio- and optical spectroscopy. It has been established, that Landau-Lifshitz equation is fundamental physical equation underlying the dynamics of spectroscopic transitions and transitional phenomena for matter subsystem of resonance interacting system field-matter. New phenomena - stochastic electrical and magnetic spin wave resonances are predicted to be the effects of EM-field quantization. 

\end{document}